\documentstyle[preprint,aps,psfig]{revtex}
\tightenlines
\begin{document}
\draft
\title{RELATIONSHIP BETWEEN DINUCLEAR SYSTEMS AND NUCLEI IN
HIGHLY DEFORMED STATES}
\author{T.M.Shneidman$^{1,2}$, G.G.Adamian$^{1,2,3}$, N.V.Antonenko$^{1,2}$,
S.P.Ivanova$^{1,2}$ and W.Scheid$^2$}
\address{$^{1}$Joint Institute for Nuclear Research, 141980 Dubna, Russia\\
$^{2}$Institut f\"ur Theoretische Physik der
Justus--Liebig--Universit\"at,
D--35392 Giessen, Germany\\
$^{3}$Institute of Nuclear Physics,
Tashkent 702132, Uzbekistan
}
\date{\today}
\maketitle

\begin{abstract}
Potential energies, moments of inertia, quadrupole and octupole moments
of  dinuclear systems are compared with the corresponding
quantities of  strongly deformed nuclei. As dinuclear system we denote
two touching nuclei (clusters).
It is found that the hyperdeformed states of nuclei
are close to those of nearly symmetric dinuclear systems, whereas
the superdeformed states are considered as states of asymmetric dinuclear
systems. The superdeformed and hyperdeformed
states constructed from two touching clusters have   large octupole
deformations. The experimental measurement of octupole deformation of
the highly deformed nuclei can answer whether these
nuclei have cluster configurations as described by the dinuclear model.
\end{abstract}

\pacs{PACS: 21.60.Ev,21.60.Gx \\ Key words:
Dinuclear system; Cluster states in nuclei; Superdeformed and hyperdeformed
states}

\section{Introduction}
One of the most important developments in nuclear structure physics was
the prediction and observation of superdeformed (SD) \cite{1} and
hyperdeformed (HD) \cite{2} nuclear shapes. The idea that nuclei could adopt
highly deformed prolate shapes at low excitation energies was originated with
the discovery of deformed isomers in the actinide region \cite{3,4}.
Another group of superdeformed states with
the ratio 3:2 of major to minor axis
was discovered near the
ground state in the $A \approx 76 $  ($^{72}$Se,$^{74,76}$Kr) \cite{5,6}
and  $A \approx 100 $  ($^{98,100}$Sr,$^{100}$Zr) \cite{7,8} mass regions.
In these nuclei  the SD ground  and  excited
states are strongly mixed with a  spherical band which
coexists at low spin \cite{9}.
While in the rare earth nuclei the highly deformed shapes are
stabilized by collective rotation, the highly
deformed nuclei exist even at zero spin.
The states with large spins are populated in heavy ion fusion reactions.
By the study of rotational bands one can determine the
moments of inertia of highly deformed nuclei. Based on the experimental
values of the moment of inertia, it was found that the SD and HD states are
related to  shapes with a ratio of
axes of 2:1 and 3:1, respectively. Since the
intensity of $\gamma$-transitions drastically decreases with
decreasing angular momentum, the
experimental determination of the excitation energy of the SD band
is difficult.

In actinides the third potential minimum of the HD state is elucidated
from a microstructure in the resonances found in the reactions ($n,f$),
($t,pf$) and ($d,pf$) \cite{10}. Another evidence supporting
the existence of a third minimum is the observation of
an asymmetric angular distribution of the light
fission fragments of nuclei around
$^{232}$Th \cite{11}. The transitions of odd multipolarity indicate
a reflection-asymmetric shape of the nucleus $^{232}$Th.

Investigations of the SD and HD rotational bands in different mass
regions were performed within semi-phenomenological cranked
Woods-Saxon and Nilsson approaches \cite{12,13}. Nuclear mean
field theories, based on these approaches, well describe a rich
variety of nuclear shapes all over the periodic table. These
models allowed predictions of HD states which were recently observed
experimentally \cite{14}. The SD and HD rotational bands have been
also investigated within relativistic mean field theories
\cite{15}.

The nuclear shapes calculated within mean field theories are close
to a rotational ellipsoid. It is known from the study of light
nuclei ($^{8}$Be,$^{32}$S) that the SD shape of nuclei can be
considered as a symmetric di-molecular shape rather than an
ellipsoid. In light $\alpha$-particle nuclei the similarity
between hyperdeformed and cluster-type states, i.e. quasi-molecular
states, was already mentioned in \cite{16,17}. Besides theoretical
work, there are experimental evidences for the existence of the
cluster-type configurations in fissioning nuclei \cite{18}. The
validity of the cluster approach for heavier nuclei has been
investigated in \cite{19,20,21,22}.
An interesting observation
in  shell model calculations is that the nucleus in the third minimum
corresponds to a dinuclear system (DNS) configuration \cite{20}.
However, in this model the clusters  penetrate  each other because
the relative distance $R$ between the centers of the clusters  is smaller
than the sum of cluster radii $R_1+R_2$.
As it was shown in \cite{23,24}, the overlapping of nuclei is
hindered by a repulsive potential
at smaller relative distances $R$. Therefore, in the present
paper we describe  the SD and HD states by molecular-like dinuclear system
configurations  with a relative distance
 $R_m\approx R_1+R_2$, which corresponds
to the minimum of the nucleus-nucleus potential \cite{17}.
In this paper we find a relationship between the DNS-type cluster
configurations and highly deformed states of heavy nuclei.
Consequences for assuming HD and SF states as
cluster-type states are discussed.

\section{Multipole moments of DNS}

The mass ($k=m$) and charge ($k=c$) multipole moments
of the DNS shape are calculated
 with the expression
\begin{eqnarray}
Q^{(k)}_{\lambda \mu}=
\sqrt{\frac{16 \pi}{2 \lambda +1}}
\int \rho^{(k)}({\bf r})r^\lambda Y_{\lambda \mu}(\Omega)d\tau.
\label{def}
\end{eqnarray}
For  small overlaps of the  nuclei in the DNS when $R \ge R_1+R_2,$ where
$R_1$ and $R_2$ are the radii of the nuclei and  $R$ the distance between
the centers of nuclei, the nuclear
mass and charge densities $\rho^{(k)}$ in the DNS
 can be written as a sum of the densities in each nucleus (frozen density
  approximation):
\begin{eqnarray}
\rho^{(k)}({\bf r})=
\rho^{(k)}_1({\bf r})+\rho^{(k)}_2({\bf r}).
\label{den}
\end{eqnarray}
By using Eq.(\ref {den}) and assuming  axial symmetry of the nuclear shapes,
the multipole moments of the DNS in the center of mass of system
can be expanded in the following form:
\begin{eqnarray}
Q^{(k)}_{\lambda 0}=Q^{(k)}_{\lambda}=
\sum^{\lambda}_{\lambda_1=0 \atop \lambda_1+\lambda_2 = \lambda }
(-1)^{\lambda}\frac{\lambda !}{\lambda_1 ! \lambda_2 !}
\left [(-1)^{\lambda_1}A_2^{\lambda_1}Q^{(k)}_{\lambda_2}(1)+
A_1^{\lambda_1}Q^{(k)}_{\lambda_2}(2) \right ]
\frac{R^{\lambda_1}}{A^{\lambda_1}},
\label{mom_dns}
\end{eqnarray}
where multipole moments of the DNS  nuclei $Q^{(k)}_{\lambda_2}(i)$ $(i=1,2)$
are calculated in their  centers of mass.
For example, up to $\lambda =3$ the values of $Q^{(k)}_{\lambda}$ are:
\begin{eqnarray}
&& Q^{(m)}_1 = 0, \nonumber \\
&& Q^{(c)}_1 = 2 e \frac{A_2 Z_1 - A_1 Z_2}{A}R, \nonumber \\
&& Q^{(m)}_2 =2 m_0  \frac{A_1 A_2}{A}R^2+
Q^{(m)}_2(1)+Q^{(m)}_2(2),  \nonumber \\
&& Q^{(c)}_2 =  2 e   \frac{A_2^2 Z_1+A_1^2 Z_2}{A^2}R^2+
Q^{(c)}_2(1)+Q^{(c)}_2(2), \nonumber \\
&& Q^{(m)}_3 =2 m_0 \frac{A_1 A_2 }{A}\frac{A_2-A_1}{A}R^3+
3 \frac{A_2Q^{(m)}_2(1)-A_1 Q^{(m)}_2(2)}{A}R, \nonumber \\
&& Q^{(c)}_3 =2 e \frac{A_2^3 Z_1-A_1^3 Z_2}{A^3}R^3+
3 \frac{A_2 Q^{(c)}_2(1)-A_1 Q^{(c)}_2(2)}{A}R.
\label{pp_mom}
\end{eqnarray}
Here, $A=A_1+A_2$ and $A_i$, $Z_i$  $(i=1,2)$
are the mass number of the system and the
mass and charge numbers of the DNS nuclei,
 respectively;  $m_0$ is the mass of the nucleon.
Experimental values of the quadrupole moments
of the  DNS nuclei are taken in the calculations.
We consider the DNS nuclei in pole-pole orientation which
corresponds to the minimum of potential energy.
Since the  diffuseness in $\rho^{(k)}_i({\bf r})$ in the nuclear
surface does not practically influence the results for the here considered
relative distances $R$, we disregard the effect of
diffuseness in the present paper.

The shape of an axially-deformed nucleus
can be described by a multipole expansion of the nuclear surface
\begin{eqnarray}
R=R_0(1+\beta_0 Y_{00} + \beta_1 Y_{10}+\beta_2 Y_{20}+\beta_3 Y_{30}),
\label{def1}
\end{eqnarray}
where $R_0$ is the spherical equivalent radius of the nucleus and
$\beta_0$, $\beta_1$, $\beta_2$, $\beta_3$ are deformation parameters with
respect to the center of mass.
The $\beta_\lambda$ are widely used to characterize the experimental
spectroscopic informations.
The parameter $\beta_0$ is responsible for preserving the nuclear
volume. The parameter $\beta_1$  provides the vanishing
dipole moment $Q^{(m)}_{10} =0$.
The deformation parameters $\beta_2$ and $\beta_3$ are related to
the quadrupole and octupole moments of the axially--deformed
nucleus. With Eqs.~(1) and (5) and assuming a constant nuclear density,
we can express the mass multipole moments through the deformation parameters
$ \beta_\lambda$ $(\lambda=0,1,2,3)$:
\begin{eqnarray}
\tilde Q_\lambda^{(m)}=\frac{3}{\lambda +3} A m_0 R^{\lambda}_{0}
\sum_{k_0=0}^{\lambda+3} \sum_{k_1=0}^{k_0}
\sum_{k_2=0}^{k_1}\sum_{k_3=0}^{k_2}
G^\lambda_{k_0 k_1 k_2 k_3} \beta_0^{k_0-k_1}
 \beta_1^{k_1-k_2} \beta_2^{k_2-k_3}\beta_3^{k_3}
\label{Qdt_eq}
\end{eqnarray}
where
\begin{eqnarray}
G^\lambda_{k_0 k_1 k_2 k_3}&=&\frac{1}{2^{\lambda +k_2}\lambda !}
\sqrt{\frac{3^{k_1-k_2}5^{k_2-k_3}7^{k_3}}{(4 \pi)^{k_0}}}
{{\lambda +3}\choose{k_0}} {{k_0}\choose{k_1}}{{k_1}\choose{k_2}}
{{k_2}\choose{k_3}}  \nonumber\\
&\times&
\sum_{i=0}^{k_2-k_3}  \sum_{j=0}^{k_3}
(-1)^{i+j}3^{k_2-k_3-i+j}5^{k_3-j} {{k_2-k_3}\choose{i}}
{{k_3}\choose{j}}I_{i j},
\nonumber
\end{eqnarray}
\begin{eqnarray}
I_{ij}=
\sum_{k=0}^{\left[\lambda/2\right]}(-1)^k {{\lambda}\choose{k}}
\frac{(2\lambda -2k)!}{(\lambda -2k)!}
\frac{2\delta_{k_1+k_2+k_3+\lambda +1,odd}}
{(k_1+k_2+k_3+\lambda +1-2i-2j-2k)}.
\nonumber
\end{eqnarray}
Here,  ${n\choose k}=n!/(k!(n-k)!)$.

In spite
of a better shape
parameterization for large deformations \cite{25} one
can approximately apply Eq.~(\ref{def1}) to the DNS.
With Eq.~(\ref{def1}) one can well describe the DNS shape
for a small mass asymmetry $|\eta|=|(A_2-A_1)/A|<0.5$.
For larger $|\eta|$, Eq.~(\ref{def1}) leads to smother shapes
than the DNS shape. However, even asymmetric configurations
can be effectively characterized by the parameters $\beta_\lambda$.
Using the experimental mass quadrupole and octupole
moments, the experimentalists obtain
the deformation parameters $ \beta_\lambda$  $(\lambda=2,3) $
with expressions equal or similar to (\ref{Qdt_eq}).
We do the same procedure but take the DNS multipole moments
$ Q^{(m)}_\lambda $
from (\ref{pp_mom}) instead of  moments of SD and HD nuclei,
extracted from  experiment. Therefore, we solve the system of equations
\begin{eqnarray}
\tilde{Q}^{(m)}_\lambda=Q^{(m)}_{\lambda}
\label{syseq}
\end{eqnarray}
and find the dependences of $\beta_\lambda$
on the mass (charge) asymmetry $\eta=(A_2-A_1)/A$ $(\eta_Z=(Z_1-Z_2)/Z)$
and the relative distance between the centers of nuclei.
 In Fig.~1 we show the dependence of the
 parameters $\beta_1$, $\beta_2$ and $\beta_3$
 on the mass asymmetry for the case that the nuclei
 of the DNS are assumed spherical.
  This dependence  is independent of the total mass number which is simply
  demonstrated if we take $A_1=A(1-\eta)/2$ and $A_2=A(1+\eta)/2$.
A dinuclear system with  spherical and  equal mass
fragments (with radii $R_0$) in touching leads visually to
the axis ratio 2:1
because it is the ratio of the DNS length (4$R_0$)
to the DNS width (2$R_0$).
However, the quadrupole moment $Q^{(m)}_2$ for this system is equal
to the quadrupole moment of ellipsoid with axis ratio 2.65:1.
With deformed nuclei the ratio will be near 3:1 as
for hyperdeformed nucleus.

The values of quadrupole and octupole parameters of deformation become
close to each other at large mass asymmetry.
For very asymmetric DNS consisting of spherical nuclei,
we can use simple analytical expressions for
$\beta_2$ and $\beta_3$:
\begin{eqnarray}
\beta_2 =\sqrt{\frac{5}{4 \pi}}\frac{4 \pi}{3}
\frac{A_1 A_2}{A^2}\frac{R^2}{R^2_0}, \nonumber \\
\beta_3 =\sqrt{\frac{7}{4 \pi}}\frac{4 \pi}{3}
\frac{A_1 A_2}{A^2}\frac{A_2-A_1}{A}
\frac{R^3}{R^3_0}.
\label{lma}
\end{eqnarray}

\section{Moment of inertia of DNS}
The moment of inertia
is usually calculated microscopically
by using the cranking-type formula and taking the
residual interactions (for example, the pairing correlations)
into consideration.
However, one can consider simpler way
to find the moment of inertia.
The overlap volume of the touching
nuclei in the DNS is about a few percent of the total DNS volume and
the individuality of the DNS nuclei is conserved. Therefore,
one can write down
the potential energy $U$ as the sum of binding energies of two nuclei
and energy of their interaction (see Sect.~4) and express
the multipole moments of
the DNS through the multipole moments
of the DNS nuclei. Due to the same reason, the DNS moment of inertia
$\Im$ can be calculated as
\begin{eqnarray}
\Im=\Im_1+\Im_2+m_0 \frac{A_1A_2}{A}R_m^2,
\label{Mi_eq}
\end{eqnarray}
where
the values of $\Im_i$ ($i$=1,2) are  the moment of inertia
of the DNS clusters and in the case of small angular momentum
are extracted
from the experimental value of the energies
$E_{2^+\to 0^+}$ of  2$^+\to$ 0$^+$ transitions:
$\Im_i=3/E_{2^+\to 0^+}$ ($\hbar^2/$MeV) \cite{26}.

For large angular momenta, the moments of inertia $\Im_i$ $(i=1,2)$
of the DNS nuclei can be with good accuracy calculated
in the rigid body approximation
\begin{eqnarray}
\Im_i&=&\frac{1}{5}m_0 A_i(a_i^2+b_i^2), \nonumber \\
a_i&=&R_{0i}(1-\frac{\alpha_i^2}{4 \pi})
(1+\sqrt{\frac{5}{4 \pi}}\alpha_i), \nonumber \\
b_i&=&R_{0i}(1-\frac{\alpha_i^2}{4 \pi})(1-\sqrt{\frac{5}{16 \pi}}\alpha_i)
\end{eqnarray}
because the experimental data \cite{Duda}
show that the moment of inertia of superdeformed states is very close
to the rigid body limit (about 85 percent of last one).
Here,  $R_{0i}$ and  $ \alpha_i$ $(i=1,2) $ are the spherical equivalent
radii and the parameters of quadrupole deformation of the DNS
nuclei, respectively.
The moment of inertia is  well measured
for the SD and HD states. Therefore, a
comparison of calculated and experimental values
of the moment of inertia can prove our interpretation
of the shapes of highly deformed nuclei.
Of course, the complete clarification of our cluster approach and the problem
how to calculate the moment of inertia could come from
attempt to compute actual spectra for direct comparison with data,
especially the low-lying negative parity bands symptomatic
of octupole deformations.  In the forthcoming paper we hope
to explain the parity splitting as the effect of tunneling along mass
asymmetry (by the analogy with tunneling along $\beta_3$ \cite{JLS})
from the asymmetic DNS with positive $\eta$ to the DNS
with the same but negative $\eta$. For this purpose, we
need more accurate
calculations of the potential energy and, correspondingly,
of the moment of inertia as a function of $L$.

\section{Potential energy of DNS}
In order to check the possibility of the DNS formation
in  the excited compound nucleus, the potential energy of the DNS
is calculated as \cite{17}
\begin{eqnarray}
U(R, \eta, L)= B_1+ B_2+  V(R,\eta,L)- [B_{12}+V^{'}_{rot}(L)].
\label{7_eq}
\end{eqnarray}
Here, $B_1$, $B_2$, and $B_{12}$ are the realistic binding
energies of the fragments and the compound nucleus,
respectively. The shell effects are included in these binding energies
and supply the local minima in $U(R, \eta, L)$ as a function of $\eta$.
The value of $U(R,\eta,L)$ is normalized to the energy
of the rotating compound nucleus by $B_{12}+V^{'}_{rot}$.  The
nucleus-nucleus potential  (\ref{7_eq})
\begin{eqnarray}
V(R,\eta,L)=V_{coul}(R,\eta) + V_N(R,\eta) + V_{rot}(R,\eta,L)
\end{eqnarray}
is the sum of the Coulomb \cite{Wong}
\begin{eqnarray}
V_{coul}=\frac{Z_1Z_2 e^2}{R}+\frac{3}{5}
\frac{Z_1Z_2e^2}{R^3}\sum_{i=1}^2 R^2_{0i}\alpha_i Y_{20}(\cos{\theta_i})+
\frac{12}{35}\frac{Z_1Z_2e^2}{R^3}
\sum_{i=1}^2 R^2_{0i}(\alpha_iY_{20}(\cos{\theta_i}))^2,
\label{coul}
\end{eqnarray}
the centrifugal \cite{17}
\begin{eqnarray}
V_{rot}=\frac{\hbar^2 L(L+1)}{2\Im}
\label{rot}
\end{eqnarray}
and the nuclear
\begin{eqnarray}
V_N(R,\eta)=\int \rho_1({\bf r}_1)\rho_2({\bf R}-{\bf r}_2)F({\bf r}_1-{\bf
r}_2)d {\bf r}_1 d {\bf r}_2
\label{fold1}
\end{eqnarray}
potentials. For the nuclear part of $V(R,\eta,L)$ we use
a double folding formalism with the Skyrme-type effective
density-dependent nucleon-nucleon interaction
\begin{eqnarray}
&& F({\bf r}_1-{\bf r}_2) = C_0 \left (
F_{in} \frac{\rho_1({\bf r_1})+\rho_2({\bf R}-{\bf r_2})}{\rho_{0}}+
F_{ex}(1-\frac{\rho_1({\bf r_1})+\rho_2({\bf R}-{\bf r_2})}{\rho_{0}})
\right ) \delta({\bf r}_1-{\bf r}_2), \nonumber \\
&& F_{in,ex}=f_{in,ex}+f^\prime_{in,ex}\frac{N_1-Z_1}{A_1}\frac{N_2-Z_2}{A_2},
\label{effpot}
\end{eqnarray}
which is well known from the theory of finite Fermi systems
\cite{17}. Here,  $N_i$ ($i$=1,2) are neutron numbers of the nuclei.
The values of $C_0$=300 MeV fm$^3$
and the dimensionless parameters $f_{in}$=0.09,  $f_{ex}$=-2.59,
$f^\prime_{in}$=0.42
and $f^\prime_{ex}$=0.54 are fitted to
describe a large number of experimental data \cite{Mig}.
For the density of heavy nuclei one can use the two-parameter
symmetrized Woods-Saxon function
\begin{eqnarray}
\rho_i({\bf r})=\frac{\rho_{0}
\sinh(R_i(\theta_i,\phi_i)/a_{i})}
{ \cosh(R_i(\theta_i,\phi_i)/a_{i})+ \cosh(r/a_{i})},
\label{SWS}
\end{eqnarray}
where $\rho_{0}=$0.17 fm$^{-3}$
is  density in the center of nucleus.
We use a nuclear radius parameter $r_0=1.14$ fm
and a diffuseness parameter $a_i=(0.54-0.55$ fm depending on the
mass number of the isotope.
For light nuclei with mass numbers
$A_i<$16, the more realistic functional dependence
\begin{eqnarray}
\rho_i({\bf r})=A_i(\gamma_i^2/\pi)^{3/2}\exp{(-\gamma_i^2r^2)}
\label{Gau}
\end{eqnarray}
is used, where $\gamma_i$ characterizes the width of the nucleon
distribution in the nucleus. The value of $\gamma_i$ can be obtained
by minimizing the nuclear binding energy in the density function \cite{17}.
The deformation effects are taken into account in the calculation
of the potential energy surface \cite{17}.
The shapes of the DNS nuclei with quadrupole deformations
can be written as
\begin{eqnarray}
R_i(\theta_i,\phi_i)=R_{0i}
(1-\frac{\alpha_i^2}{4 \pi}+\alpha_i Y_{20}(\theta_i,\phi_i)).
\label{shape}
\end{eqnarray}

With the nucleon-nucleon interaction (\ref{effpot})
the repulsive core appears in $V_N(R,\eta)$ which prevents
the motion to smaller distances  ($R<R_1+R_2$) and reflects
the action of the Pauli principle.
For $R_1+R_2-1$ fm $<R<R_1+R_2+2$ fm the potential $V(R,\eta,L)$
as a function of the relative distance $R$
has a pocket with small depth
which is resulted by the atractive nuclear and repulsive Coulomb
interactions.
The DNS is localized in the minimum of
this pocket at $R_m\approx R_1+R_2
=R_{01}(1+\sqrt{\frac{5}{4 \pi}} \alpha_1)+
R_{02}(1+\sqrt{\frac{5}{4 \pi}} \alpha_2)$ for the pole-pole configuration.
The relative orientation of the deformed nuclei in the DNS follows
the minimum of the potential energy which yields the pole-pole
orientation.
At fixed mass asymmetry $\eta$, charge asymmetry $\eta_Z$
and deformation parameters $\alpha_i$ ($i$=1,2)
the minimum of  $V(R,\eta,L)$ of the DNS
corresponds to the distance $R=R_m(\eta,\eta_Z,\alpha_i)$
when the poles of nuclei
touch each other.
For each $\eta$, we minimize $U(R_m,\eta,L)$ with respect to
the charge asymmetry. Since the minimization of $U$ with respect
to $\alpha_i$ ia very cumbersome, we
took from Refs.~[26,27] the parameters of
the quadrupole deformation which are
related to the excitation of the first $2^+$ states
if the energies of these states are
smaller than 1.5 MeV.
As is known from experiments on sub-barrier fusion, these states
are easily populated.
For the treatment of small internal excitation of the DNS
nuclei, this approximation seems to be good for deriving
the minimum of $U$ with respect to $\alpha_i$.
So, the energy $U(R_m,\eta,L)$ plotted as a function of
$\eta$ results minimum values $B_1+B_2+V(R,\eta,L)$
with respect to other degrees of freedom in the DNS.

The potentials for all DNS
were calculated with the same set of  parameters
and assumptions. We found that the final results are not crucial to
a reasonable variation of these parameters in the calculation
of the potential energy.
Since the  excitation energies of the DNS nuclei are relatively
small, the calculation of the DNS potential
energy immediately shows which DNS configuration can be related to
the SD or HD nuclear systems.
Among all minima in $U(R_m,\eta,L)$ as a function of $\eta$
we selected those with the minimal values of $U$.
These minima appear for small and large $\eta$ and we
analysed either almost symmetric or very asymmetric
configurations. If there are several minima with
close energies in some interval of $\eta$, all of them are
treated. However, the calculated values of $Q_2$ and $\Im$ would
be similar for these DNS.

\section{Results and discussions}
With the formalism given in Sections II-IV we studied various
nuclei described with the dinuclear system concept. As compound
nuclei we chose $^{152}$Dy, $^{232}$Th, $^{234}$U, $^{240}$Pu,
$^{76}$Kr, $^{148}$Nd and $^{236}$Ra which will be discussed next.

a) $^{152}$Dy:

The dependences of $\beta_2$, $\beta_3$, $Q_2$ ($Q_2=Q^{(c)}_2$)
and $Q_3$ ($Q_3=Q^{(c)}_3$)
on the mass asymmetry $\eta$ are presented in Fig.~2
for the DNS corresponding to the $^{152}$Dy compound nucleus.
Since the deformations of the DNS nuclei are functions of $\eta$,
these dependences have some oscillations.
While $Q_2$ and $\beta_2$ decrease with increasing $\eta$,
 $Q_3$ and $\beta_3$  have
maxima. The positions of these maxima approximately correspond
to  maximal potential energy of the DNS as a function of $\eta$ (see Fig.~3).
The value of $\beta_3$  increases quickly with
$\eta$ from zero.  For very asymmetric DNS,
$\beta_3$ becomes again relatively small.
Therefore, a given value of $\beta_3$ can correspond
to two different DNS with different mass asymmetries. For the symmetrical
DNS with spherical nuclei,
the difference by a factor $\sqrt{2}$ between our value of $\beta_2$
and $\beta_2$ obtained in \cite{22} is due to another
definition of $\beta_2$ in \cite{22}.

The DNS potential energy  as a function
of the mass asymmetry is shown in Fig.~3 for  the cases of spherical
and deformed nuclei in the DNS forming $^{152}$Dy.
The calculation with  deformed nuclei
 yields various  minima of the potential energy, especially
  at $\eta=0.026$ ($^{74}$Ge+$^{78}$Se), $\eta=0.16$ ($^{64}$Ni+$^{88}$Sr) and
$\eta=0.34$ ($^{50}$Ti+$^{102}$Ru). For zero angular momentum, the
energy of the combination $^{50}$Ti+$^{102}$Ru is about 20 MeV
which is close to the value estimated for the HD state in
\cite{28}. We calculated the DNS moment of inertia $\Im$ using
the experimental moments $\Im_i$ of the DNS nuclei ($\Im=\Im^e$) or
the rigid body moments $\Im_i$ given in eq.(10) ($\Im=\Im^r$). The
first value of the moment of inertia of the system
$^{50}$Ti+$^{102}$Ru is obtained as $\Im^e=100 \hbar^2$MeV$^{-1}$
and the second one as $\Im^r=131 \hbar^2$MeV$^{-1}$. The latter
value is close to the experimental one $\Im^{exp}=130\hbar^2$MeV$^{-1}$ \cite{2}.
For this DNS, the obtained value $\beta_2=1.3$ is in agreement
with the experimental estimate $\beta_2 \ge 0.9$. Therefore, we conclude
that the shape of the DNS $^{50}$Ti+$^{102}$Ru is compatible with the
shape of the $^{152}$Dy nucleus in the HD state.

We found that in the asymmetric DNS, for example
$^{22}$Ne+$^{130}$Ba ($\eta$=0.71) and $^{26}$Mg+$^{126}$Xe
($\eta$=0.66) (where the potential energy has minima),
the calculated moments of inertia and quadrupole
moments are close to the experimental values $\Im^{exp}=(85\pm 3)\hbar^2\,
$MeV$^{-1}$ and $Q_2^{exp}=(18\pm 3)e\,$ barn,
respectively, known for the SD state.
For the system $^{22}$Ne+$^{130}$Ba, we calculated
$\Im^r=100 \hbar^2\,$MeV$^{-1}$ (and $\Im^{e}=63\hbar^2\,$MeV$^{-1}$),
$Q_2=20 e\,$ barn and $\beta_2=0.8$.
In the system $^{26}$Mg+$^{126}$Xe we obtained
$\Im^r=104 \hbar^2\,$MeV$^{-1}$ (and $\Im^e=67 \hbar^2\,$MeV$^{-1}$),
 $Q_2=24 e\,$ barn and $\beta_2=0.9$. These DNS  have practically zero
  temperature if they are formed in the reaction $^{48}$Ca(205 MeV)
  ($^{108}$Pd,4n)$^{152}$Dy. At $L=0$ the potential energies of these DNS
with respect to the ground state of $^{152}$Dy lie about 8 MeV
higher than those estimated for the SD shapes in \cite{28}.
The energies of these two systems with respect
to the energy of the compound nucleus are shown in Fig.~4
as a function of the square of the angular momentum.

The SD bands in the mass  region $A \approx 150$ are populated with
an anomalously high intensity around a spin of $55\hbar$.
There is no significant population of these states  below
$(45-50) \hbar$ \cite{29}. This fact can be explained by the DNS
interpretation of the SD states.
The dependence of the DNS potential energy on $\eta$
becomes flat with decreasing angular momentum and the DNS
can evolve to larger $\eta$.
This takes place because the potential barrier in the direction of
larger mass asymmetries decreases with $L$ (Fig.~3).
Due to the distribution of the excitation energy among
many configurations at
small $L$ and due to the motion of the DNS to the
compound nucleus with increasing
$\eta$, the transition from the SD state to the ground state is invisible.

b) $^{232}$Th:

Analysing the potential energy of the DNS as a function
of $\eta$ for the $^{232}$Th compound nucleus (see Fig.~5),
we found well exposed minima corresponding
to the systems $^{100}$Zr+$^{132}$Sn ($\eta=0.138$) and
$^{82}$Ge+$^{150}$Ce ($\eta=0.293$).  Fig.~6 shows  the  dependence
of $\beta_2$ and $\Im=\Im^r$ on $\eta$ which is
weak for small mass asymmetries.
  In both $^{232}$Th systems,  $\beta_2$ is about 1.5,
$\beta_3\approx 0.40$,  $\Im^r$ about $290 \hbar^2\,$MeV$^{-1}$
and $\Im^e \approx 240 \hbar^2\,$MeV$^{-1}$;
the potential  energies, given in Fig.~5, are near the energy of the ground
state of $^{232}$Th. Excepting the values of $\beta_2$,
the obtained values in Table~1 are close to the
corresponding ones calculated for the third
minimum in \cite{20}. In our calculation $\beta_2$ is
larger in comparison to $\beta_2=0.85$ in \cite{20}.
The SD rotational bands in $^{232}$Th can be interpreted as the
DNS states of the configurations
$^{28}$Mg$+^{204}$Pt and $^{26}$Ne$+^{206}$Hg (Table~1).

c) $^{234}$U and $^{240}$Pu:

For the $^{234}$U nucleus, we found  the SD and HD
cluster configurations (Table~1):
$^{26}$Ne$+^{208}$Pb,$^{28}$Mg$+^{206}$Hg,$^{82}$Ge$+^{152}$Nd,
$^{100}$Zr$+^{134}$Te and $^{104}$Mo$+^{130}$Sn.
The experimentally extracted depth of the third well (which corresponds
to the HD state) was determined to be $(3.6 \pm 0.3)$ MeV for the $^{234}$U
nucleus \cite{30}.
For the DNS configurations,
this value is an agreement with the value of the depth of
the pocket in the nucleus-nucleus potential as function of $\eta$.
For example, for the
$^{100}$Zr$+^{134}$Te and $^{104}$Mo$+^{130}$Sn configurations, the
depth of the pocket is about 3.4 MeV.

Fig.~5 shows also the potential energy of the $^{240}$Pu compound
nucleus where SD bands were observed in an experiment on
sub-barrier fission \cite{30}.
These SD states could be considered as the DNS-type states:
$^{32}$Mg$+^{208}$Pb and $^{34}$Si$+^{206}$Hg (see Table~1).
From the results presented in Table 1 one can predict some HD cluster states
for the $^{240}$Pu nucleus.

d) $^{74}$Kr and $^{72}$Se:

The potential energy of the DNS for the $^{76}$Kr compound
nucleus is shown in Fig.~5.
We find a deep minimum with an energy on the level
of the energy of the compound nucleus for the cluster configuration
$^{8}$Be+$^{68}$Ge. For this cluster state we have
$Q_2$ =4.9 $e$~barn, $Q_3=2.0\times 10^3$ $e$~fm$^3$ and
$\Im^r=29 \hbar^2$/MeV ($\Im^e=17.2 \hbar^2$/MeV). A similar
 picture is observed in other nuclei in the $A \approx 76$ mass region,
for example,  in $^{74}$Kr and  $^{72}$Se.
 In the DNS  configuration with  the  light
 cluster $^{8}$Be  we found $Q_2=4.7$ $e$~barn,
 $Q_3=1.9\times 10^3$ $e$~fm$^3$ for the
 $^{72}$Se compound nucleus. For the nuclei in the mass region
 $A \approx 100$, the energies of such
cluster configurations  are by 5 to 6 MeV higher
than the energies of corresponding compound nuclei.

Let us consider the  $N/Z$-equilibrium. In the nuclei $^{72}$Se,$^{74,76}$Kr
we obtained minima in  the DNS potential energy for  configurations
with $\alpha$-particle nuclei (multiple of $\alpha$-clusters) at large $\eta$.
The potential energies of these DNS are small because the
light nuclei in them are most stable.
Such behaviour of the DNS potential energy is not observed for  asymmetric
configurations with light clusters in the $A \approx 100$ mass region.

e) $\alpha$-structure and dipole and octupole deformations,
$^{148}$Nd, $^{226}$Ra and $^{220-228}$Th:

If very asymmetric DNS are energetically
favorable, the wave function of the compound nucleus has components
belonging to  cluster-type configurations.
For example, in many cases the cluster configurations
\begin{eqnarray}
^{A}Z \to ^{(A-4)}(Z-2)+^4 {\rm He}
\label{he}
\end{eqnarray}
have an energy which is close or even lower than the energy of the ground
state of the compound nucleus. As a result, the
nucleus can have an octupole deformation in the ground state.
Since the shape of the nucleus is no more symmetric under space
inversion, the spectrum of such a nucleus must contain  states with different
parity. This was experimentally found in many nuclei with
$Z \approx 88-90$ $(N \approx 86-90)$, in different isotopes of the nuclei
Ra, Th and U, and with $Z \approx 60$, in isotopes of
Ba, Ce, Nd, Sm and Gd \cite{31}.
Another consequence of the asymmetric shape is the appearance  of
E1 and E3 transitions. These transitions were found
in $^{226}$Ra with $Q_2$=750 $e$~fm$^2$,
$Q_3$=3100 $e$~fm$^3$ \cite{33} and in
$^{148}$Nd with $Q_2$=400 $e$~fm$^2$, $Q_3$=1500 $e$~fm$^3$ \cite{34}.
The experimental dipole moments of this nuclei are found to be
0.16 $e$~fm for $^{226}$Ra \cite{33} and 0.32 $e$~fm
for $^{148}$Nd \cite{34}.
Under the assumption that the cluster configuration (\ref{he})
mainly contributes to the ground state with a static octupole
deformation, we can obtain  the  values of the multipole
moments for these nuclei.
For $^{226}$Ra, we found $Q_1=Q_1^{(c)}/2$=4 $e$~fm, $Q_2$=776 $e$~fm$^2$,
$Q_3$=2662 $e$~fm$^3$.
For  $^{148}$Nd, we obtained $Q_1$=3 $e$~fm,
$Q_2$=486 $e$~fm$^2$, $Q_3$=1844 $e$~fm$^3$.
The calculated values of $Q_2$ and $Q_3$ are close to the experimental
ones mentioned before.
However, the values of $Q_1$ are larger by at least one
order of magnitude. The same problem was observed
in the cluster model \cite{32}.
These deviations of the theoretical dipole moments
from the experimental ones
are due to the use of a simplified treatment
of the $N_1/Z_1$-ratios in the DNS nuclei.
The value of $Q_1$ strongly depends on these ratios
and vanishes ($Q_1=0$) in the
limit of the same $N/Z$-ratio in the DNS nuclei.
For  very asymmetric DNS, the $N/Z$-ratios in the
light nucleus are effectively larger than the  $N/Z$-ratio in the
$\alpha$-particle ($N/Z=1$)because the heavy nucleus of the DNS is strongly
overlapping with the $\alpha$-particle and there is at least one
valence neutron supplying the coupling of the  $\alpha$-particle with
the heavy cluster.
If we take  the ($^4$He+1$n$) cluster instead of $^4$He in the DNS or
slightly increase the $N/Z$-ratio in the contact region of the two nuclei
of the DNS, then the theoretical value of $Q_1$ for $^{226}$Ra
and $^{148}$Nd results in  agreement with
the experimental data, but with $Q_2$ and $Q_3$ practically not changed.
For the nuclei with octupole deformation in the ground state,
one can try to explain the parity splitting as a function of $L$
by the tunneling in $\eta$ in the DNS with $\alpha$-particle.

As shown in Fig.~7,  the octupole (quadrupole) deformation becomes smaller
 (larger) with an increasing   atomic number of the Th isotopes in the
 systems $^{(A-4)}$Ra$+^4$He. This can be explained by the  fact that  the
 quadrupole deformation of the
 heavy cluster of the DNS gets  larger in  this mass region.
 Such a behaviour of $\beta_2$ and $\beta_3$ is in  agreement with the one
 obtained in  \cite{35}.

\section{Summary}
The DNS potential energy  as a function of mass asymmetry
has a few global minima. Most of them lie higher in energy than
the energy of the compound nucleus.
It is possible, however, to populate
these states in heavy ion induced reactions by choosing appropriate
reaction partners with an appropriate  bombarding energy.
At  high spins, these cluster states can be cold and have  long lifetime.
We found the energies,  moments of inertia and  quadrupole
deformations  of  certain DNS  close to the experimental ones of SD
and HD nuclei. Since many DNS states have relatively
large octupole deformations,
the experimental measurement of the octupole deformation of highly deformed
nuclei can answer the question whether these nuclei exist
in  cluster configurations.
An evidence, that SD nuclei can have octupole deformations, is for example
 an observation of an excited SD band in $^{190}$Hg which decays to the
lowest-energy (yrast) SD-band by  transitions of odd multipolarity.
The E1 rate observed in \cite{36} is three orders of magnitude
larger then those rates typically observed in heavy deformed nuclei and
is similar to those observed in octupole-unstable
normally-deformed actinide nuclei \cite{37}.

In some nuclei with $ A \approx 230$ or $A \approx 76$ the potential
energy of the DNS has minima which lie roughly on the same
energy as the energy of the compound nucleus. This means that such cluster
states can exist at low spins.
We attempted to describe the nuclei with static octupole
deformation in the ground state like the DNS where the
$\alpha$ - cluster is the lighter cluster.
It was found that calculated quadrupole
and octupole deformations are close to the
experimental ones and that such DNS are
energetically favorable.

\acknowledgments
We thank Prof. R.V.Jolos for fruitful discussions and Yu.V.Palchikov
for valuable advices regarding the calculations.
G.G.A. and  T.M.S. are grateful  the Alexander
von Humboldt-Stiftung and the European Physical Society
for support, respectively.
This work was supported in part by DFG and RFBR.

\begin{table}
\caption{The  values of moments of inertia $\Im^r$ and $\Im^e$,
charge quadrupole moments $Q_2$ and octupole moments $Q_3$ ,
quadrupole and octupole deformation parameters $\beta_2$ and
$\beta_3$ for different DNS corresponding to the compound nuclei
$^{232}$Th, $^{234}$U, $^{240}$Pu (see text).}
\begin{tabular}{|c|c|c|c|c|c|c|}
Clust. Conf. & $\Im^r$ ($\hbar^2$/MeV) & $\Im^e$ ($\hbar^2$/MeV) &
$Q_2\times 10^2$($e$~fm$^2)$ & $Q_3\times 10^3$($e$~fm$^3)$
& $\beta_2$ & $\beta_3$ \\
\hline
$^{26}$Ne$+^{206}$Hg $\to^{232}$Th & 171 & 82& 24.9 & 18.8 & 0.57 & 0.63 \\
$^{28}$Mg$+^{204}$Pt$ \to ^{232}$Th & 180 & 94 &31.0 & 23.9 & 0.65 & 0.68 \\
$^{82}$Ge$+^{150}$Ce $\to^{232}$Th & 292 & 249 & 70.9 & 19.8 & 1.53 & 0.47 \\
$^{100}$Zr$+^{132}$Sn $\to^{232}$Th & 292 & 235 & 70.1 & 16.2 & 1.48 & 0.34 \\
\hline
$^{26}$Ne$+^{208}$Pb$\to ^{234}$U & 169 & 79 & 20.9 & 18.0 & 0.47 & 0.61 \\
$^{28}$Mg$+^{206}$Hg$\to ^{234}$U & 179 & 91 & 29.9 & 23.7 & 0.61 & 0.68 \\
$^{82}$Ge$+^{152}$Nd$\to ^{234}$U & 291 & 257 & 70.4 & 20.5 & 1.49 & 0.49 \\
$^{100}$Zr$+^{134}$Te$\to ^{234}$U & 326 & 258 &71.8 & 14.0 & 1.73 & 0.27 \\
$^{104}$Mo$+^{130}$Sn$\to ^{234}$U & 296 & 242 & 71.8 & 14.0 & 1.48 & 0.28 \\
\hline
$^{32}$Mg$+^{208}$Pb$\to ^{240}$Pu & 191 & 101 & 28.7 & 23.2 & 0.57 & 0.71 \\
$^{34}$Si$+^{206}$Hg$\to ^{240}$Pu & 197 & 107 & 33.9 & 25.4 & 0.69 & 0.70 \\
$^{82}$Ge$+^{158}$Sm$\to ^{240}$Pu & 307 & 267 & 75.3 & 22.1 & 1.53 & 0.50 \\
$^{104}$Zr$+^{136}$Xe$\to ^{240}$Pu & 305 & 253 & 74.1 & 14.8 & 1.49 & 0.33 \\
$^{106}$Mo$+^{134}$Te$\to ^{240}$Pu & 314 & 258 & 76.4 & 16.0 & 1.52 & 0.32 \\
$^{110}$Ru$+^{130}$Sn$\to ^{240}$Pu & 311 & 250 & 75.3 & 12.4 & 1.49 & 0.23 \\
\end{tabular}
\end{table}

\begin{figure}
\centerline{\psfig{figure=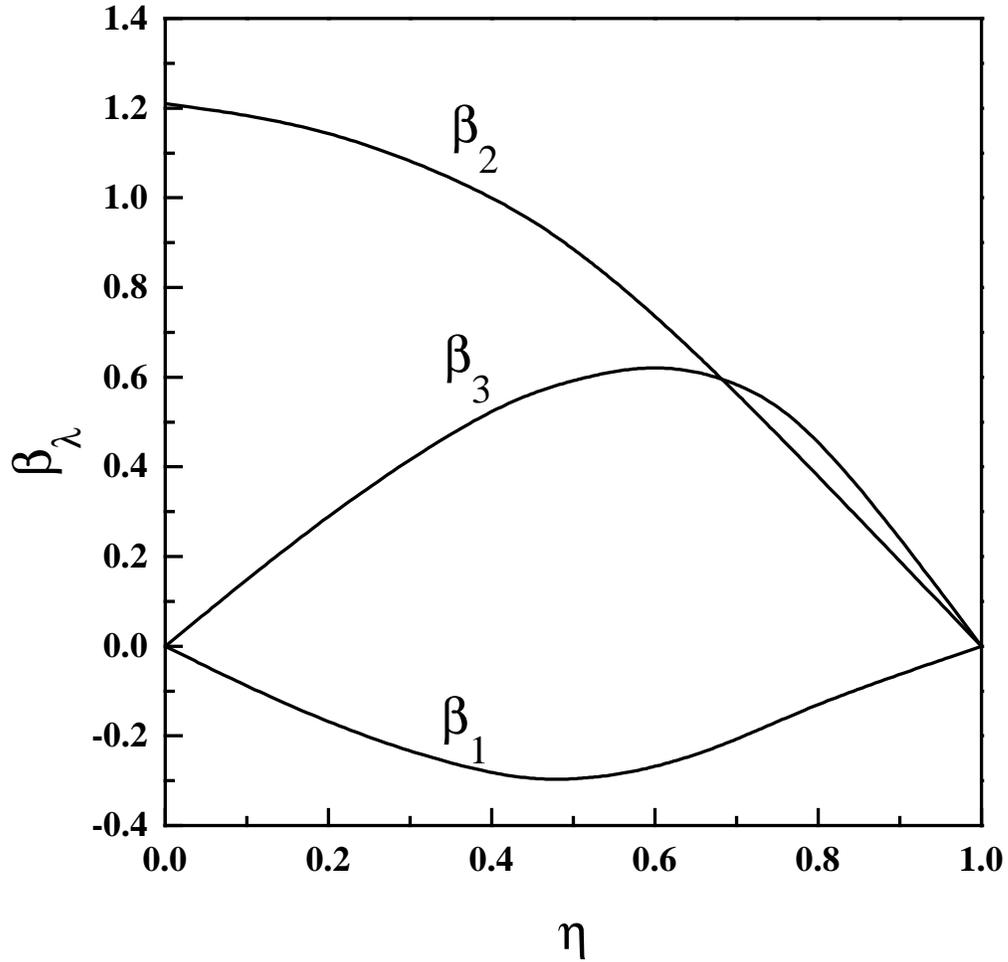,width=14.5cm}}
\caption{Dependence of the deformation parameters
$\beta_1$, $\beta_2$, $\beta_3$ on mass asymmetry
$\eta$ for a DNS with spherical nuclei. The parameters  do not
depend on the total  mass number of the DNS.}
\end{figure}
\newpage

\begin{figure}
\centerline{\psfig{figure=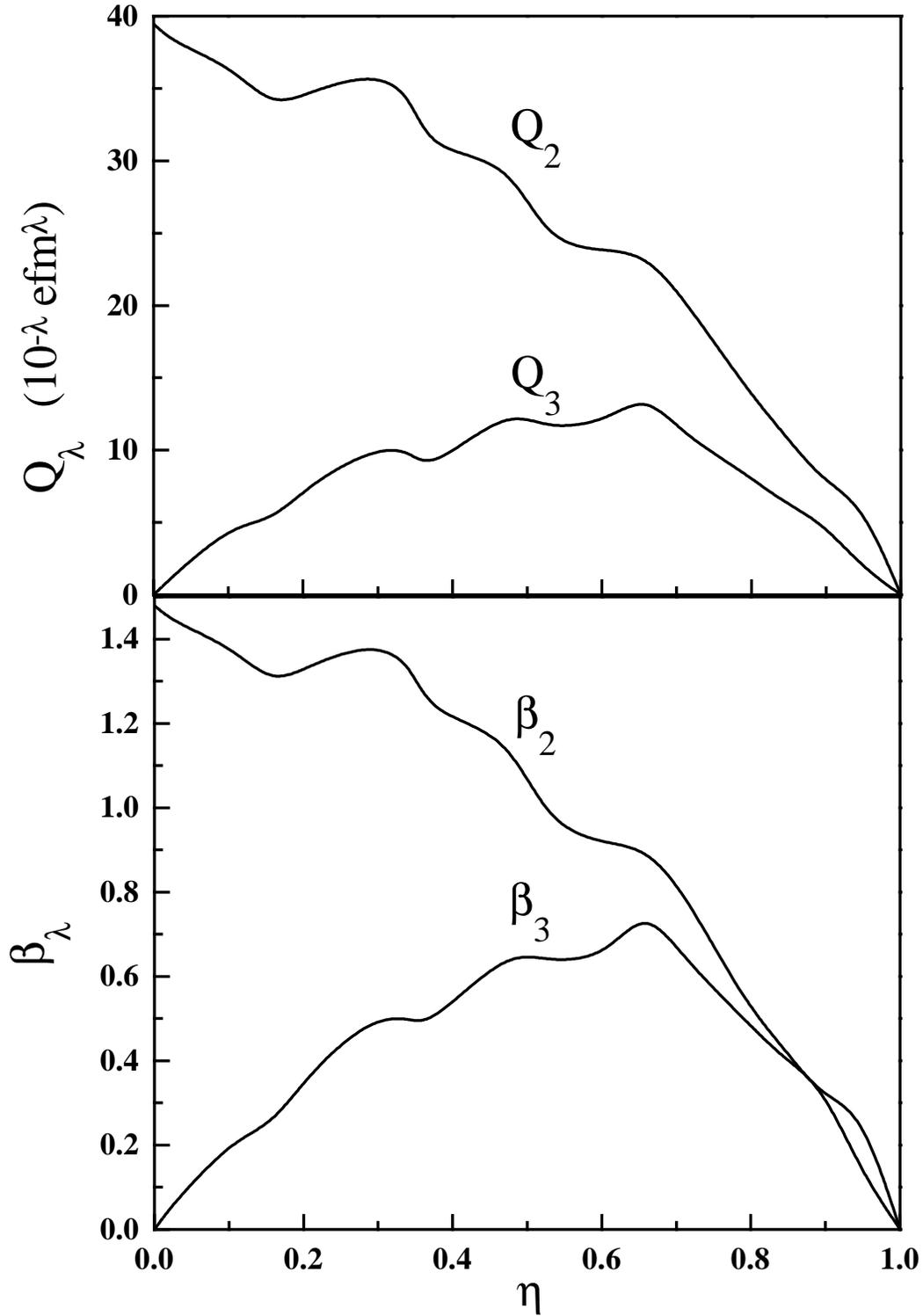,width=14.5cm}}
\caption{Dependence of $Q_2$ ($10^{-2}$ $e$~fm$^2$) and
$Q_3$ ($10^{-3}$ $e$~fm$^3$) (upper part) and of $\beta_2$ and $\beta_3$
(lower part) on the  mass asymmetry $\eta$ of the DNS for the
$^{152}$Dy compound nucleus.
The deformation of the DNS nuclei are taken into account.}
\end{figure}
\newpage

\begin{figure}
\centerline{\psfig{figure=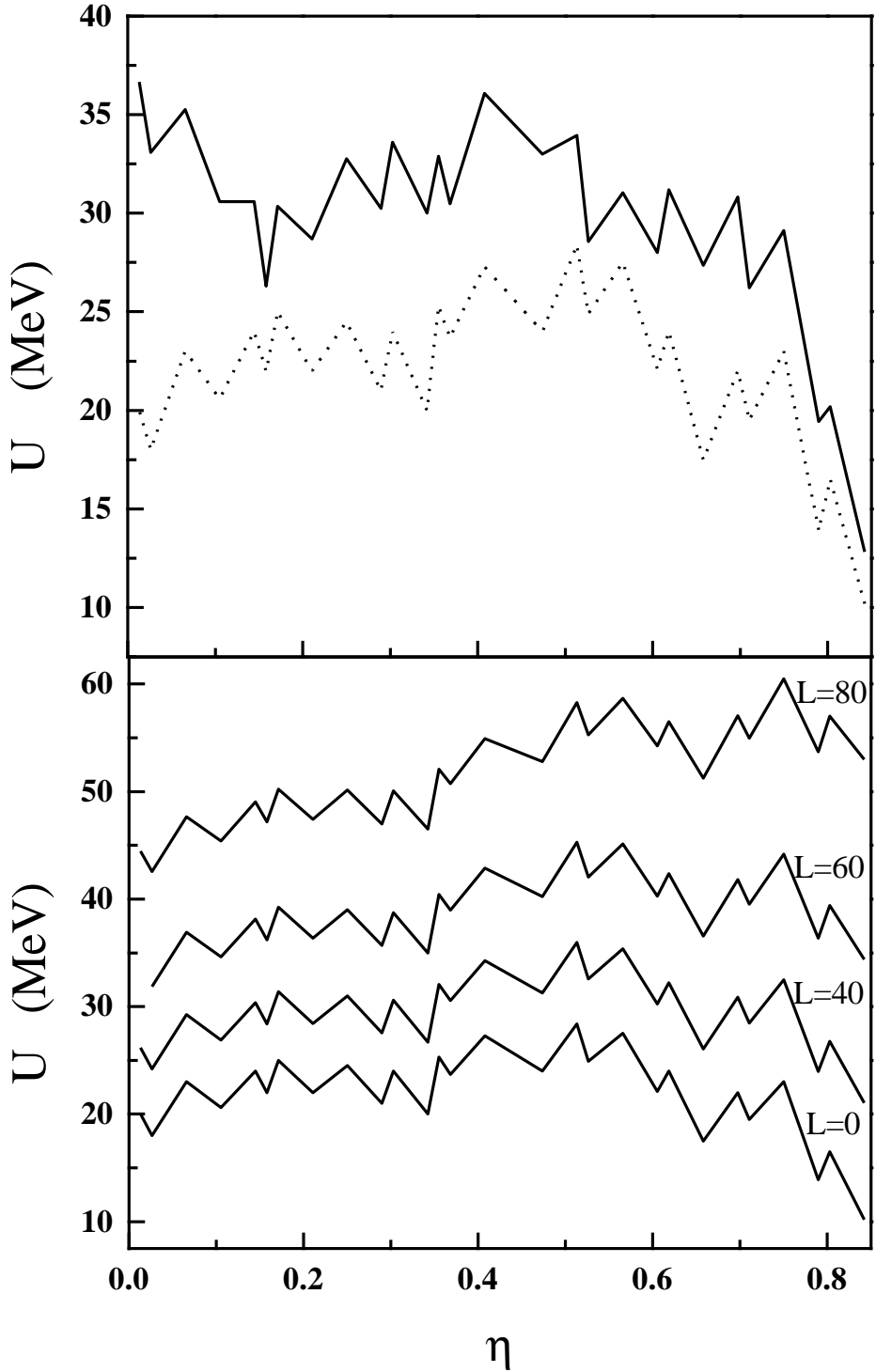,width=14cm}}
\caption{Potential energy $U$ of the DNS as a function of
mass asymmetry $\eta$ for the $^{152}$Dy compound nucleus.
At $L=0$ the calculated results without and with deformation of the
DNS nuclei are presented by solid and dotted lines, respectively
(upper part). The lower part shows
results for different  angular momentum quantum numbers
calculated  with deformed DNS nuclei.}
\end{figure}
\newpage

\begin{figure}
\centerline{\psfig{figure=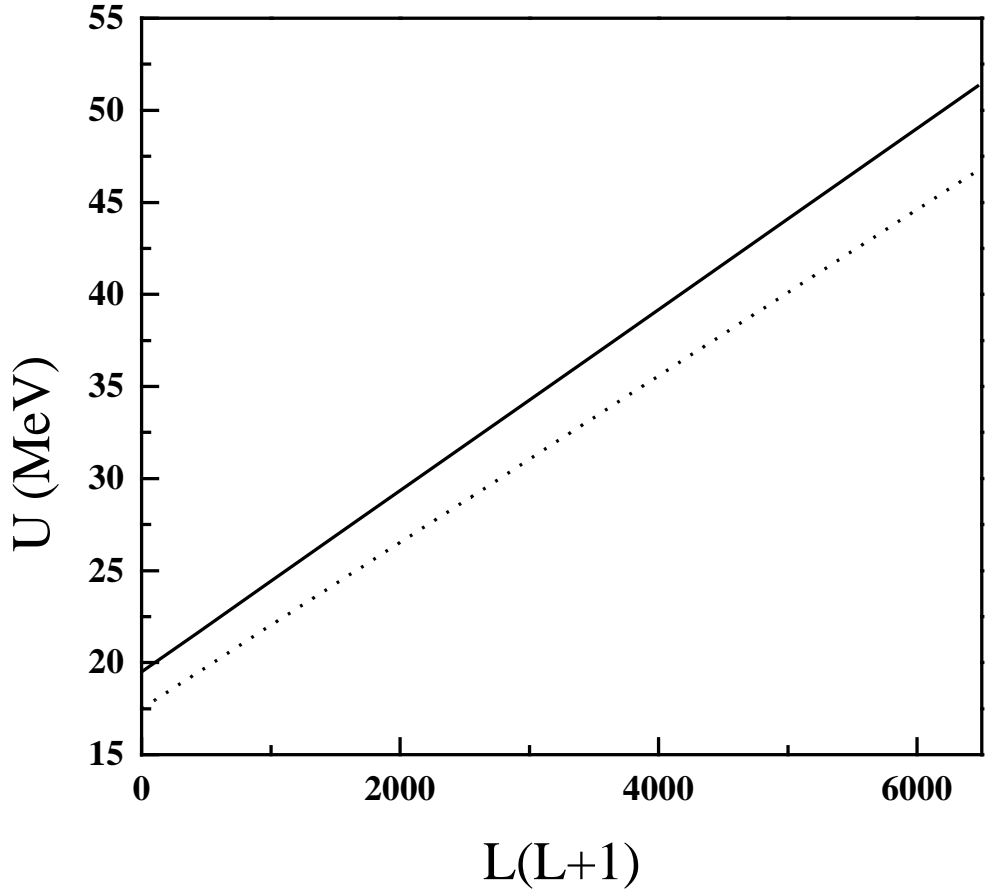,width=14.5cm}}
\caption{Dependence of the energies of the systems  $^{22}$Ne+ $^{130}$Ba
(dotted line) and  $^{26}$Mg+$^{126}$Xe (solid line)  on the
square of the angular momentum quantum number.
The energy is normalized to the energy of the rotating
compound nucleus.}
\end{figure}
\newpage

\begin{figure}
\centerline{\psfig{figure=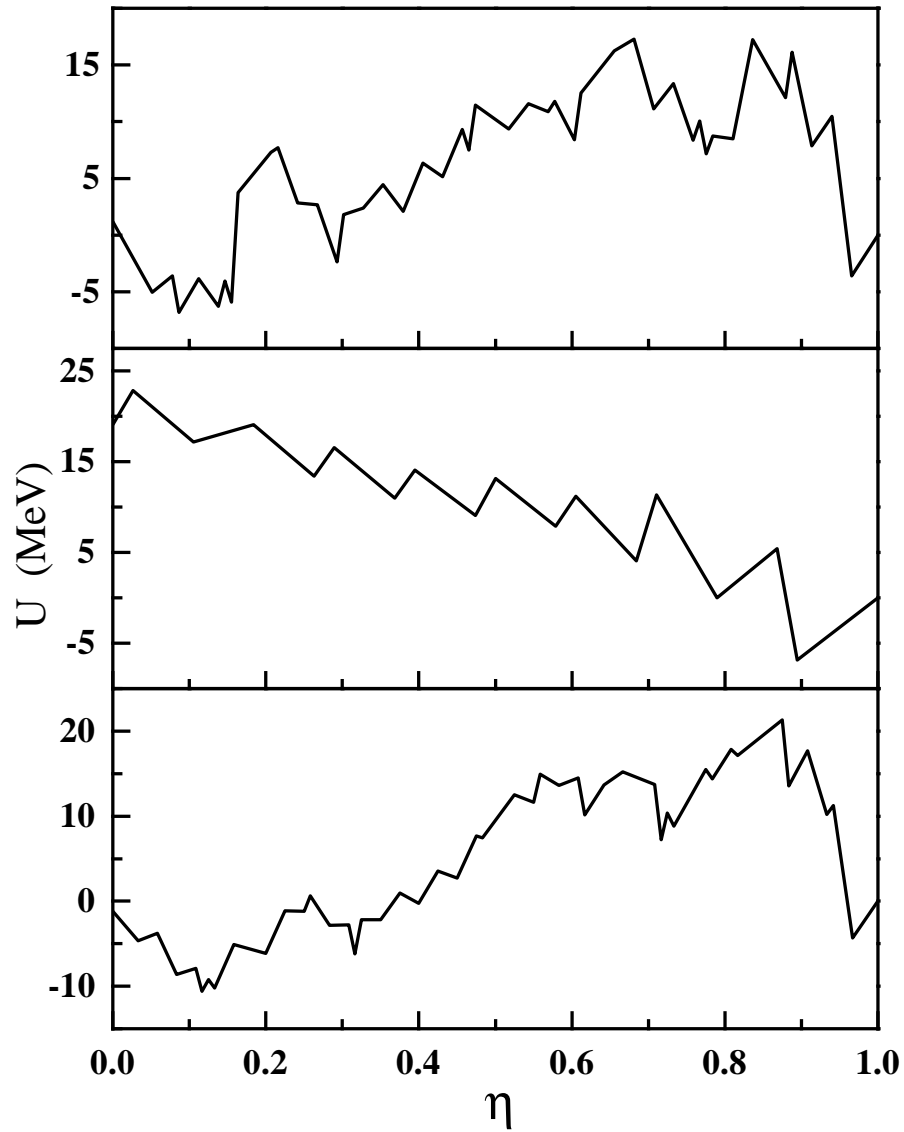,width=14.5cm}}
\caption{Potential energy $U$ of the DNS as a function of $\eta$
for the compound nuclei
$^{232}$Th (upper part), $^{76}$Kr (middle part) and $^{240}$Pu (lower part)
at $L=0$. The deformation of the DNS nuclei is taken into account.}
\end{figure}
\newpage

\begin{figure}
\centerline{\psfig{figure=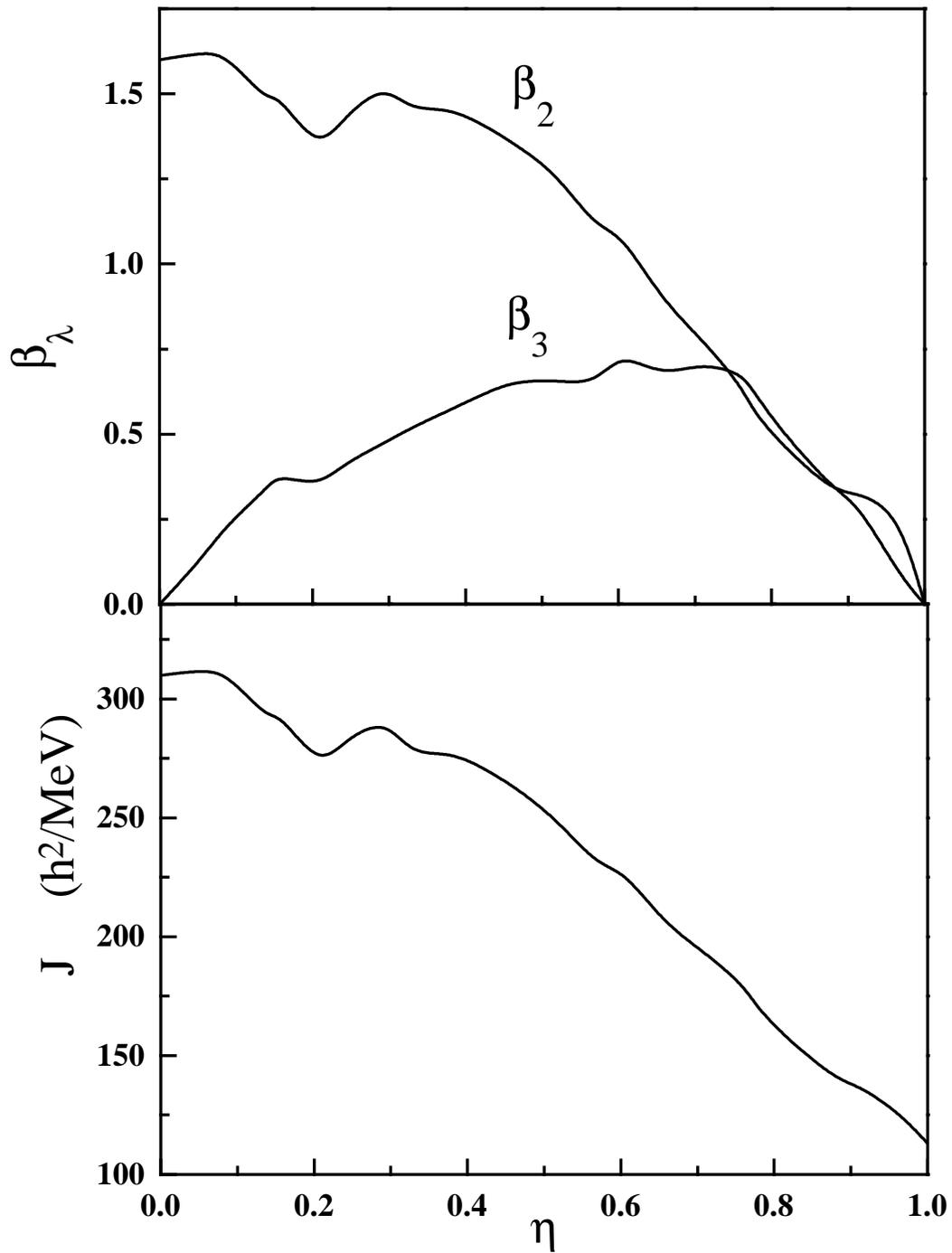,width=14.5cm}}
\caption{Deformation parameters $\beta_2$ and  $\beta_3$ (upper part)
and moment of inertia $\Im=\Im^r$ (lower part) as a function of mass
asymmetry $\eta$ of the DNS corresponding
 to the $^{232}$Th compound nucleus.}
\end{figure}
\newpage

\begin{figure}
\centerline{\psfig{figure=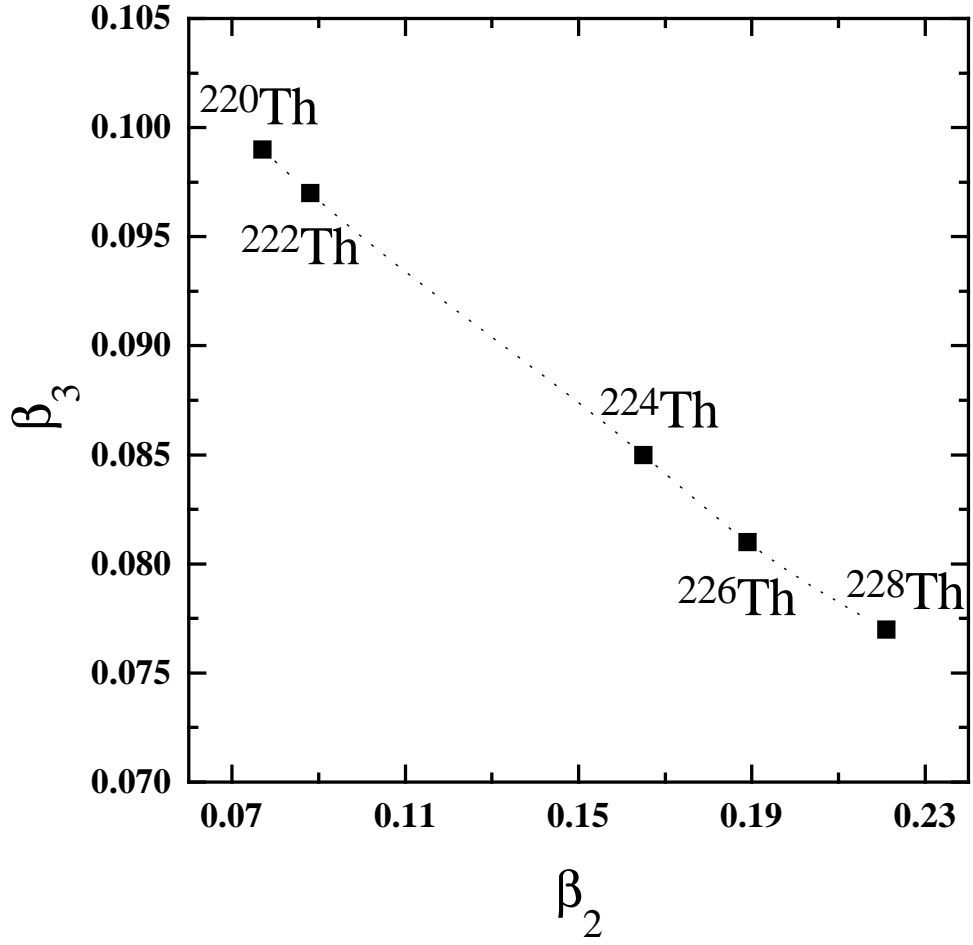,width=14.5cm}}
\caption{The
$\beta_2$ vs. $\beta_3$ plot for different isotopes of $^A$Th in the cluster
state $^{(A-4)}$Ra+$^4$He.}
\end{figure}

\end{document}